\pdfoutput=1

\documentclass[11pt]{article}

\usepackage[]{emnlp2022}

\usepackage{times}
\usepackage{latexsym}

\usepackage[T1]{fontenc}

\usepackage[utf8]{inputenc}

\usepackage{microtype}

\usepackage{inconsolata}

\usepackage{multirow}

\usepackage{amssymb}

\usepackage{mathtools}
\usepackage{placeins}
\usepackage{adjustbox}
\DeclareMathOperator*{\argmax}{argmax}
\newcommand{\amsearch}{Top-$K$-Search}
\newcommand{\amseq}{{Top-$K$-Train}}

\usepackage{cleveref}
\usepackage{todonotes}
\usepackage{csquotes}

\title{Does Joint Training Really Help Cascaded Speech Translation?}

\author{
Viet Anh Khoa Tran \; David Thulke \; Yingbo Gao \; Christian Herold \; Hermann Ney \\
Human Language Technology and Pattern Recognition Group \\
Computer Science Department\\
RWTH Aachen University \\
D-52056 Aachen, Germany \\
{\tt \{vtran|thulke|gao|herold|ney\}@i6.informatik.rwth-aachen.de}
}

\begin{document}
\maketitle
\begin{abstract}
Currently, in speech translation, the straightforward approach - cascading a recognition system with a translation system - delivers state-of-the-art results.
However, fundamental challenges such as error propagation from the automatic speech recognition system still remain.
To mitigate these problems, recently, people turn their attention to direct data and propose various joint training methods.
In this work, we seek to answer the question of whether joint training really helps cascaded speech translation.
We review recent papers on the topic and also investigate a joint training criterion by marginalizing the transcription posterior probabilities.
Our findings show that a strong cascaded baseline can diminish any improvements obtained using joint training, and we suggest alternatives to joint training.
We hope this work can serve as a refresher of the current speech translation landscape, and motivate research in finding more efficient and creative ways to utilize the direct data for speech translation.
\end{abstract}

\section{Introduction}

Speech translation (ST) is the task of automatic translation of speech in some source language into some other target language \cite{stentiford1988machine, waibel1991janus}.
Traditionally, a cascaded approach is used, where an automatic speech recognition (ASR) system is used to transcribe the speech, followed by a machine translation (MT) system, to translate the transcripts \cite{sperber-paulik-2020-speech}.
The problem of error propagation has been the center of discussion in ST literature \cite{ney1999speech, casacuberta2004some, matusov2005phrase, peitz2012spoken, sperber-etal-2017-neural}, and instead of using the discrete symbols in the source languages, ideas like using n-best lists, lattices, and neural network hidden representations are investigated \cite{saleem2004using, Kano2017StructuredBasedCL, anastasopoulos-chiang-2018-tied, zhang-etal-2019-lattice, sperber2019attention}.
For a more systematic review of the ST development, we refer the readers to \citet{sperber-paulik-2020-speech}.

With recent efforts in the expansion of the ST data collection \cite{di2019must, beilharz-etal-2020-librivoxdeen}, more and more direct ST data is available.
Such direct data comes as pairs of source speech and target translation, and often as triplets further including source transcriptions.

Various joint training methods are proposed to use such data to improve cascaded systems, with the hope that uncertainties during transcription can be passed on to translation to be resolved. Here, what we call \enquote{joint training} is often referred to as \enquote{end-to-end training} in the literature, where the direct ST data is utilized in the joint optimization of the ASR and MT models \cite{Kano2017StructuredBasedCL, Berard2018EndtoEndAS, anastasopoulos-chiang-2018-tied, inaguma2019multilingual, sperber2019attention, bahar-etal-2019-using, wang2020bridging, baharTightIntegratedEndtoEnd2021}.
In this work, we revisit the principal question of whether or not joint training really helps cascaded speech translation.

\section{Cascaded Approach}
In traditional cascaded systems, an ASR model $p(f_1^J|x_1^T)$ and an MT model $p(e_1^I|f_1^J)$ are trained separately, where we denote speech features as $x_1^T$, transcriptions as $f_1^J$, and translations as $e_1^I$.
The decoding is done in two steps:
\begin{align*}
    \hat{f}_1^J&=\argmax_{[f_1^J]} p(f_1^J|x_1^T)\\
      \hat{e}_1^I
      &=\argmax_{[e_1^I]} p(e_1^I\mid\hat{f}_1^J)
    \end{align*}
The $\argmax$ is approximated using beam search for computational reasons, and we will assume a fixed beam size $N$ for the decoding of both transcriptions and translations.

\section{Joint Training Approaches}

\subsection{Top-$K$ Cascaded Translation}

Assume we have pre-trained an ASR and an MT model, and some direct ST training data is available.
The pre-trained ASR model is used to produce a $K$-best list of ASR hypotheses $F_1,F_2,\ldots, F_K$ using beam search with beam size $N\ge K$.
While there is no unique method to make use of the top-$K$ transcript, we describe \textbf{\amseq{}}, a straightforward algorithm similar to re-ranking.
We obtain the score $\tilde{p}$ for each ASR hypothesis with length normalization and normalize them locally within the top-$K$ hypotheses.
\begin{equation}\label{eq:asr_score}
  p(F_k|x_1^T) = \frac{\tilde{p}(F_k|x_1^T)}{\sum_{k'=1}^{K}\tilde{p}(F_{k'}|x_1^T)}
\end{equation}
During training, $p(e_i|F_K;e_0^{i-1})$ is the MT model output.
Given the ASR hypotheses $F_1,\ldots,F_K$, the following training objective is maximized.
\begin{equation*}
  \log\left(
    \sum_{k=1}^{K} p(F_k|x_1^T) \prod_{i=1}^{I} p(e_i|F_k;e_0^{i-1})  
  \right)
\end{equation*}
We hypothesize that this objective (a) exposes different transcriptions and potential ASR errors to the MT model and (b) encourages the ASR model to produce hypotheses closer to the expectations of the MT model, thus reducing model discrepancy.
Since discrete ASR hypotheses are passed to the MT model from a previous beam search, the error signal to ASR is passed via the renormalized transcript scores during backpropagation.

Similarly, we introduce \textbf{\amsearch{}}. We obtain an MT hypothesis $E_k$ for each $F_k$ using beam search. The final hypothesis is obtained as below.
\begin{equation*}
  \hat{e}_1^{\hat{I}} = \argmax_{E_k}\,\{p(F_k|x_1^T)\cdot p(E_k|F_k)\}
\end{equation*}
Here, $p(F_k|x_1^T)$ is obtained as in Equation~\ref{eq:asr_score} and $p(E_k|F_k)$ is the length normalized translation score from the MT model.
Observe that this search is applicable to any cascade architecture and is thus independent of the training criterion. 
In our experiments, we always \amsearch{} when decoding models trained with \amseq{}.
The idea of generating the top-$K$ ASR hypotheses during search has also been explored in the literature (e.g.\ Section~\ref{sec:searchable}).

\subsection{Tight-Integration}
Another way to train the cascade architecture using direct ST data is the \textbf{tight integrated cascade} approach \cite{baharTightIntegratedEndtoEnd2021}.
We introduce an exponent $\gamma$ that controls the sharpness of the distribution of the conditional probabilities. Thus, instead of passing the 1-best hypothesis of the ASR system as a sequence of 1-hot vectors, we pass 
the renormalized probabilities to the MT model.
\begin{equation*}
  p(f_j|f_1^{j-1}; x_1^T) = \frac{\tilde{p}^{\gamma}(f_j|f_1^{j-1}x_1^T)}{\sum_{f'\in |V_F|}\tilde{p}^{\gamma}(f'_j|f_1^{j-1}x_1^T)}
\end{equation*}
Here, $V_F$ is the vocabulary of the ASR system.

\subsection{Searchable Hidden Intermediates}\label{sec:searchable}

\citet{dalmia2021searchable} propose passing 
the final decoder representations
of the $N$-best ASR hypotheses (i.e.\ the \textbf{searchable hidden intermediates}) directly to the MT system, bypassing the MT input embedding.

Additionally, they extend the multi-task learning approach by allowing the MT decoder to attend to the ASR encoder states, which in turn are optimized using beam search in training.
They show that during decoding, a higher ASR beam size indeed leads to a better ST performance.

\section{Experimental Results and Analysis}
\newcommand{\dataminus}{-}
\begin{table*}[!ht]
\centering
\begin{tabular}{ll|ccc|cccc}
\hline
   & \multirow{3}{*}{\textbf{Model}}                   & \multicolumn{3}{c|}{\textbf{Data}}     & \multicolumn{4}{c}{\textbf{MuST-C En-De}}      \\ \cline{3-9} 
 &
   &
  \multirow{2}{*}{\textbf{MT}} &
  \multirow{2}{*}{\textbf{ASR}} &
  \multirow{2}{*}{\textbf{ST}} &
  \multicolumn{2}{c|}{\textbf{tst-HE}} &
  \multicolumn{2}{c}{\textbf{tst-COMMON}} \\ \cline{6-9} 
 &
   &
   &
   &
   &
  \multicolumn{1}{c}{\textbf{BLEU}} &
  \multicolumn{1}{c|}{\textbf{TER}} &
  \multicolumn{1}{c}{\textbf{BLEU}} &
  \multicolumn{1}{c}{\textbf{TER}} \\ \hline
   & \citet{baharTightIntegratedEndtoEnd2021}          &  &    &  &      & \multicolumn{1}{c|}{}     &      &      \\
   & cascade                                           & \checkmark & \checkmark   & \dataminus{}  & 25.0 & \multicolumn{1}{c|}{59.2} & 25.9 & 56.2 \\
   & tight integrated cascade                          & \checkmark & \checkmark   & \checkmark & 26.8 & \multicolumn{1}{c|}{57.2} & 26.5 & 54.8 \\ \hline
   & \citet{xu2021stacked}                             &  &   & &      & \multicolumn{1}{c|}{}     &      &      \\
   & cascade                                           & \dataminus{}  & \dataminus{}    & \checkmark & -    & \multicolumn{1}{c|}{-}    & 23.3 & -    \\
   & cascade                                           & \checkmark & \checkmark   & \checkmark & -    & \multicolumn{1}{c|}{-}    & 28.1 & -    \\
   & direct + pretraining + multi-task                 & \dataminus{}  & \dataminus{}    & \checkmark & -    & \multicolumn{1}{c|}{-}    & 25.2 & -    \\
   & direct + pretraining + multi-task                 & \checkmark & \checkmark   & \checkmark & -    & \multicolumn{1}{c|}{-}    & 28.1 & -    \\ \hline
   & \citet{dalmia2021searchable}                      &   &  &   &      & \multicolumn{1}{c|}{}     &      &      \\
   & searchable hidden intermediates                   & \dataminus{}  & \dataminus{}    & \checkmark & -    & \multicolumn{1}{c|}{-}    & 26.4 & -    \\ \hline
   & \citet{inagumaESPnetSTIWSLT20212021}              &   &   & &      & \multicolumn{1}{c|}{}     &      &      \\
   & cascade                                           & \checkmark & \checkmark   & -          & 26.1 & \multicolumn{1}{c|}{-}    & 29.4 & -    \\
   & direct + KD                                       & \checkmark & \checkmark   & \checkmark & 27.4 & \multicolumn{1}{c|}{-}    & 30.9 & -    \\
   & searchable hidden intermediates + KD              & \checkmark & \checkmark   & \checkmark & -    & \multicolumn{1}{c|}{-}    & 30.8 & -    \\ \hline
   & \textbf{this work (no fine-tuning)}               &   &     &  &      & \multicolumn{1}{c|}{}     &      &      \\
A1 & ground-truth transcript MT                        & \checkmark & (\checkmark) & \dataminus{}  & 30.4 & \multicolumn{1}{c|}{54.0} & 32.5 & 48.9 \\
A2 & cascade                                           & \checkmark & \checkmark   & \dataminus{}  & 27.7 & \multicolumn{1}{c|}{57.8} & 29.0 & 53.6 \\
A3 & tight integrated cascade                          & \checkmark & \checkmark   & \checkmark & 28.7 & \multicolumn{1}{c|}{56.1} & 29.2 & 53.5 \\
A4 & \amseq                                            & \checkmark & \checkmark   & \checkmark & 28.7 & \multicolumn{1}{c|}{56.9} & 30.0 & 52.5 \\
A5 & \amsearch                                         & \checkmark & \checkmark   &            & 28.2 & \multicolumn{1}{c|}{57.1} & 29.4 & 53.1 \\
   & \textbf{this work (fine-tuned ASR and MT)} &  & &  &      & \multicolumn{1}{c|}{}     &      &      \\
B1 & ground-truth transcript MT                        & \checkmark & (\checkmark) & \dataminus{}  & 31.1 & \multicolumn{1}{c|}{53.2} & 33.7 & 48.2 \\
B2 & cascade                                           & \checkmark & \checkmark   & \dataminus{}  & 29.1 & \multicolumn{1}{c|}{56.6} & 30.5 & 52.2 \\
B3 & tight integrated cascade                          & \checkmark & \checkmark   & \checkmark & 29.4 & \multicolumn{1}{c|}{55.9} & 30.1 & 52.5 \\
B4 & \amseq                                            & \checkmark & \checkmark   & \checkmark & 29.4 & \multicolumn{1}{c|}{55.9} & 30.5 & 51.9 \\
B5 & \amsearch                                         & \checkmark & \checkmark   & \dataminus{} & 29.4 & \multicolumn{1}{c|}{56.4} & 30.8 & 51.9 \\ \hline
\end{tabular}%
\caption{Results measured in BLEU {[}\%{]} and TER {[}\%{]} on the MuST-C En-De task. Fine-tuning refers to additional phase of training exclusively on the in-domain subset of the training data of both ASR and MT models.}
\label{tab:main_results}
\end{table*}

We focus on the MuST-C English-German speech translation task \cite{di2019must} in the domain of TED talks and evaluate on test-HE and test-COMMON.
We use an in-house filtered subset of the IWSLT 2021 English-German dataset as in~\citet{baharTightIntegratedEndtoEnd2021}, which contains 1.9M segments (2300 hours) of ASR data and 24M parallel sentences of MT data. 
The in-domain ASR data comprises MUST-C, TED-LIUM, and IWSLT TED, while the out-of-domain ASR data consists of EuroParl, How2, LibriSpeech, and Mozilla Common Voice.
For translation, the dataset contains 24M parallel sentences of in-domain translation data (MuST-C, TED-LIUM, and IWSLT TED), as well as out-of-domain translation data (NewsCommentary, EuroParl, WikiTitles, ParaCrawl, CommonCrawl, Rapid, OpenSubtitles2018).
For ST data, we only use MuST-C.
We provide further details in Appendix \ref{sec:appendix_setup}.
Depending on whether or not fine-tuned on in-domain ASR and MT data, we split our experiments into two sets: A1-A5 and B1-B5.

Without fine-tuning, we observe that both the tight integrated cascade \cite{baharTightIntegratedEndtoEnd2021} (A3) and our marginalization approach (A4) outperform the baseline cascade (A2) on both test-HE and test-COMMON (Table~\ref{tab:main_results}).
However, after fine-tuning both ASR and MT models, we do not observe significant improvements of the joint training methods (B3, B4) over the cascade baseline (B2) anymore.

Our experimental results suggest that the joint training of cascaded speech translation models does not seem to be effective.
This poses the questions: why is that and what were we trying to achieve with joint training anyways?
\citet{sperber-paulik-2020-speech} highlighted three main issues with ST, and in the following, we will discuss these issues from the perspective of joint training.

\paragraph{Mismatched spoken and written domains}
Transcripts and translation data usually differ in e.g. linguistic style and punctuation. This mismatch poses a challenge for cascaded models, as translation models may struggle to handle transcript-style text.
As \citet{sperber-paulik-2020-speech} point out, some of the issues such as differences in punctuation can already be tackled by plain text normalization.

More generally, one can fine-tune the models on in-domain transcript-like ASR, MT, and ST data.
It is unusual to find ST datasets that do not come with corresponding ASR and MT data, as ST data acquisition usually involves translating from transcriptions.
Thus, we can simply fine-tune the ASR and MT models on these in-domain datasets.

Our results suggest that fine-tuning the ASR and MT models is comparable or even superior to fine-tuning these models in an end-to-end fashion on the respective speech translation dataset.
However, \citet{inagumaESPnetSTIWSLT20212021} and \citet{baharTightIntegratedEndtoEnd2021} report significant improvements of their joint cascaded approach, which is similar to the tight integrated cascade, over their cascaded baseline (Table~\ref{tab:main_results}).

What is the reason for this disparity? We pin it down to one major difference: the use of in-domain ST data, or more precisely, the lack thereof. 
\citet{inagumaESPnetSTIWSLT20212021} report that by fine-tuning on MuST-C and ST-TED, they are unable to significantly improve their MT baseline, and thus, the MT component in their cascade model is not fine-tuned on the domain of TED talks.
In contrast, we find that in-domain fine-tuning significantly improves our cascaded model, especially if applied to both the ASR and MT models (Table~\ref{tab:cascaded-ft}), also improving the individual components, as we observe a decrease in WER {[}\%{]} from 9.3 to 8.1 of the ASR component and an increase in BLEU {[}\%{]} from 32.5 to 33.7 of the MT component on tst-COMMON.
As pointed out earlier, fine-tuning diminishes any improvement obtained using any of the joint training methods we implement, as the cascade baseline significantly gains in performance.

Thus, in-domain fine-tuning is essential in order to tackle the disparity between the spoken and the written domain for vanilla cascaded models.
This especially holds true for the MT model, which is trained on non-transcript-like data, but we want it to adapt to transcript-like inputs and transcript-like outputs (with punctuations, casing, etc.).

Intuitively, instead of in-domain fine-tuning, a simple remedy is to only use in-domain data for ASR and MT.
\citet{xu2021stacked} observe an improvement of their multi-task learning method when allowing only MuST-C data (Table~\ref{tab:main_results}). 
However, the improvement vanishes as they introduce external ASR and MT data.
Since the latter represents a more realistic data scenario, we believe that the inclusion of external ASR and MT data is necessary to obtain meaningful results.

In our fine-tuning setup, we do not only adapt to transcript-like data, but also to the specific domain of TED talks.
In the literature, ST performance is commonly evaluated on test sets in the same domain as the ST data, e.g. TED talks (i.e.\ MuST-C or IWSLT TED), LibriSpeech \cite{kocabiyikoglu-etal-2018-augmenting} and CallHome \cite{post-etal-2013-improved}.
However, this poorly reflects real-life data conditions, because large amounts of in-domain ST data are not always available, while in-domain ASR and MT data is more accessible.
As a consequence, we believe that end-to-end models are artificially favored over cascade models in these setups.
We thus motivate future research to also consider out-of-domain or general-domain ST datasets, while allowing in-domain ASR and MT data.

\begin{table}[]
\centering
\resizebox{\linewidth}{!}{%
\begin{tabular}{l|cc|cc}
\hline
                                           & \multicolumn{2}{c|}{\textbf{Fine-tuning}} & \multicolumn{2}{c}{\textbf{MuST-C En-De}} \\ \cline{2-5} 
                                           & \textbf{ASR}        & \textbf{MT}         & \textbf{tst-HE}   & \textbf{tst-COMMON}   \\ \hline
\small\citet{inagumaESPnetSTIWSLT20212021} & -                   & -                   & 26.1              & 29.4                  \\ \hline
\multirow{3}{*}{\small\textbf{Our work}}   & -                   & -                   & 27.7              & 29.0                  \\
                                           & $\checkmark$        & -                   & 27.9              & 29.5                  \\
                                           & $\checkmark$        & $\checkmark$        & 29.1              & 30.5                  \\ \hline
\end{tabular}%
}
\caption{Performance of our cascaded system under different in-domain fine-tuning conditions, results measured in BLEU [\%].}
\label{tab:cascaded-ft}
\end{table}

\paragraph{Error propagation}
In case the ASR produces an error in the transcript or intermediate representations, this error is propagated to the translation model, which does not have any knowledge about the transcription process.
\citet{sperber-paulik-2020-speech} discuss this phenomenon under the term \textit{erroneous early decisions}.

Intuitively, a remedy for this issue is joint training, as we allow the MT component to learn to use information that is missing or erroneous in the intermediate representation.
For example, the tight integrated approach~\cite{baharTightIntegratedEndtoEnd2021} addresses this issue by expressing uncertainties in the transcription as posterior probabilities, while \citet{dalmia2021searchable} propose speech attention, i.e.\ a Transformer cross-attention sub-module in the MT component, attending over ASR encoder representations.

However, we make the case that joint training is not necessarily the only remedy to error propagation.
Therefore, we consider a cheating experiment based on \amsearch. For each of the top-4 ASR hypotheses, we pick the translations generated by the MT model based on the sentence-level BLEU with the ground-truth target. 
In these experiments, we obtain a BLEU {[}\%{]} score of 32.4 on tst-HE and 34.2 on tst-COMMON, in both cases outperforming the oracle MT using ground-truth transcripts.
Thus, on average, the translation of one of the top-4 transcripts generated by the ASR model is no worse than using the ground-truth transcript (B1).
Hence, we posit that error propagation can be alleviated by plain ASR re-ranking.
A possible starting point is our \amsearch{} (A5, B5), giving small, but consistent improvements over their respective cascade baselines (A2, B2) without any joint training.

Similarly, instead of sequence-level reranking using the joint ASR-MT score, \citet{dalmia2021searchable} propose augmenting the token-level ASR probabilities with either the CTC scores from the ASR encoder or an external LM score during beam search to incorporate ASR uncertainties.

\paragraph{Information loss}
Information loss occurs as the ASR model does not pass on auditory information such as intonation and timing, which may be relevant for the translation component.
While we have no empirical evidence on the significance of such information on the final performance, we observe that our MT model, given the ground-truth transcripts, still significantly outperforms any speech translation model we investigate, doing so without any auditory information. 
Furthermore, our cheating experiments suggest one may even improve over ground-truth transcriptions without any auditory information.
We posit that as of now, focusing on closing that gap is of higher importance.

\section{Conclusion}
In this work, we analyzed several reasons why joint training does not appear to help when individual automatic speech recognition and machine translation components are stronger.
We point out that cascaded models achieve state-of-the-art performance when fine-tuned on in-domain ST data.

While we do not suggest that joint training is not worth exploring, we want to encourage future research to consider data conditions more carefully and produce strong cascade baselines.
Concretely, we suggest (1) the inclusion of external ASR and MT data, (2) training strong cascade baselines by fine-tuning both ASR and MT models on in-domain transcript-like data, if available and (3) the investigation of data conditions where only out-of-domain ST data is available, while allowing in-domain ASR and MT data.

\section*{Limitations}
In our experiments, we focus only on the MuST-C En-De task due to computational constraints.
Further experiments on different language pairs could e.g.\ show differences of how spoken-written domain mismatch affects different languages.
Again, experiments in different domains without direct ST data could further underline or refute our conclusions.

In our analysis, we include experimental results from other authors as we do not have the resources to reproduce every method.
It is possible that differences in data filtering, data preprocessing, architectural choices, etc.\ could affect the comparability of these results.

We have only analyzed a subset of the joint cascade methods described in literature. 
A systematic overview of such methods is outside the scope of our work.

In order to be comparable to other works in literature, we mostly draw our conclusions using BLEU and TER.
We acknowledge that using other automatic evaluation metrics and making use of human evaluation would strengthen the significance of our findings.

\section*{Acknowledgements}
This work was partially supported by the project HYKIST funded by the German Federal Ministry of Health on the basis of a decision of the German Federal Parliament (Bundestag) under funding ID ZMVI1-2520DAT04A, and by NeuroSys which, as part of the initiative \enquote{Clusters4Future}, is funded by the Federal Ministry of Education and Research BMBF (03ZU1106DA).

We thank Parnia Bahar for sharing her data preprocessing and training recipes.

\bibliography{zotero_khoa}

\begin{thebibliography}{34}
\expandafter\ifx\csname natexlab\endcsname\relax\def\natexlab#1{#1}\fi

\bibitem[{Anastasopoulos and Chiang(2018)}]{anastasopoulos-chiang-2018-tied}
Antonios Anastasopoulos and David Chiang. 2018.
\newblock \href {https://doi.org/10.18653/v1/N18-1008} {Tied multitask learning
  for neural speech translation}.
\newblock In \emph{Proceedings of the 2018 Conference of the North {A}merican
  Chapter of the Association for Computational Linguistics: Human Language
  Technologies, Volume 1 (Long Papers)}, pages 82--91, New Orleans, Louisiana.
  Association for Computational Linguistics.

\bibitem[{Bahar et~al.(2021)Bahar, Bieschke, Schl{\"u}ter, and
  Ney}]{baharTightIntegratedEndtoEnd2021}
Parnia Bahar, Tobias Bieschke, Ralf Schl{\"u}ter, and Hermann Ney. 2021.
\newblock \href {https://doi.org/10.1109/SLT48900.2021.9383462} {Tight
  {{Integrated End-to-End Training}} for {{Cascaded Speech Translation}}}.
\newblock In \emph{2021 {{IEEE Spoken Language Technology Workshop}}
  ({{SLT}})}, pages 950--957.

\bibitem[{Bahar et~al.(2019)Bahar, Zeyer, Schl{\"u}ter, and
  Ney}]{bahar-etal-2019-using}
Parnia Bahar, Albert Zeyer, Ralf Schl{\"u}ter, and Hermann Ney. 2019.
\newblock \href {https://aclanthology.org/2019.iwslt-1.22} {On using
  {S}pec{A}ugment for end-to-end speech translation}.
\newblock In \emph{Proceedings of the 16th International Conference on Spoken
  Language Translation}, Hong Kong. Association for Computational Linguistics.

\bibitem[{Beilharz et~al.(2020)Beilharz, Sun, Karimova, and
  Riezler}]{beilharz-etal-2020-librivoxdeen}
Benjamin Beilharz, Xin Sun, Sariya Karimova, and Stefan Riezler. 2020.
\newblock \href {https://aclanthology.org/2020.lrec-1.441}
  {{L}ibri{V}ox{D}e{E}n: A corpus for {G}erman-to-{E}nglish speech translation
  and {G}erman speech recognition}.
\newblock In \emph{Proceedings of the 12th Language Resources and Evaluation
  Conference}, pages 3590--3594, Marseille, France. European Language Resources
  Association.

\bibitem[{Berard et~al.(2018)Berard, Besacier, Kocabiyikoglu, and
  Pietquin}]{Berard2018EndtoEndAS}
Alexandre Berard, Laurent Besacier, Ali~Can Kocabiyikoglu, and Olivier
  Pietquin. 2018.
\newblock End-to-end automatic speech translation of audiobooks.
\newblock \emph{2018 IEEE International Conference on Acoustics, Speech and
  Signal Processing (ICASSP)}, pages 6224--6228.

\bibitem[{Casacuberta et~al.(2004)Casacuberta, Ney, Och, Vidal, Vilar,
  Barrachina, Garc{\i}a-Varea, Llorens, Mart{\i}nez, Molau
  et~al.}]{casacuberta2004some}
Francisco Casacuberta, Hermann Ney, Franz~Josef Och, Enrique Vidal, Juan~Miguel
  Vilar, Sergio Barrachina, Ismael Garc{\i}a-Varea, David Llorens, C{\'e}sar
  Mart{\i}nez, Sirko Molau, et~al. 2004.
\newblock Some approaches to statistical and finite-state speech-to-speech
  translation.
\newblock \emph{Computer Speech \& Language}, 18(1):25--47.

\bibitem[{Dalmia et~al.(2021)Dalmia, Yan, Raunak, Metze, and
  Watanabe}]{dalmia2021searchable}
Siddharth Dalmia, Brian Yan, Vikas Raunak, Florian Metze, and Shinji Watanabe.
  2021.
\newblock Searchable hidden intermediates for end-to-end models of decomposable
  sequence tasks.
\newblock In \emph{Proceedings of the 2021 Conference of the North American
  Chapter of the Association for Computational Linguistics: Human Language
  Technologies}, pages 1882--1896.

\bibitem[{Di~Gangi et~al.(2019)Di~Gangi, Cattoni, Bentivogli, Negri, and
  Turchi}]{di2019must}
Mattia~A Di~Gangi, Roldano Cattoni, Luisa Bentivogli, Matteo Negri, and Marco
  Turchi. 2019.
\newblock Must-c: a multilingual speech translation corpus.
\newblock In \emph{2019 Conference of the North American Chapter of the
  Association for Computational Linguistics: Human Language Technologies},
  pages 2012--2017. Association for Computational Linguistics.

\bibitem[{Inaguma et~al.(2019)Inaguma, Duh, Kawahara, and
  Watanabe}]{inaguma2019multilingual}
Hirofumi Inaguma, Kevin Duh, Tatsuya Kawahara, and Shinji Watanabe. 2019.
\newblock \href {https://doi.org/10.1109/ASRU46091.2019.9003832} {Multilingual
  end-to-end speech translation}.
\newblock In \emph{2019 IEEE Automatic Speech Recognition and Understanding
  Workshop (ASRU)}, pages 570--577.

\bibitem[{Inaguma et~al.(2021)Inaguma, Yan, Dalmia, Guo, Shi, Duh, and
  Watanabe}]{inagumaESPnetSTIWSLT20212021}
Hirofumi Inaguma, Brian Yan, Siddharth Dalmia, Pengcheng Guo, Jiatong Shi,
  Kevin Duh, and Shinji Watanabe. 2021.
\newblock Espnet-st iwslt 2021 offline speech translation system.
\newblock \emph{IWSLT 2021}, page 100.

\bibitem[{Kano et~al.(2017)Kano, Sakti, and
  Nakamura}]{Kano2017StructuredBasedCL}
Takatomo Kano, Sakriani Sakti, and Satoshi Nakamura. 2017.
\newblock Structured-based curriculum learning for end-to-end english-japanese
  speech translation.
\newblock In \emph{INTERSPEECH}.

\bibitem[{Kim et~al.(2017)Kim, Hori, and Watanabe}]{kim2017joint}
Suyoun Kim, Takaaki Hori, and Shinji Watanabe. 2017.
\newblock Joint ctc-attention based end-to-end speech recognition using
  multi-task learning.
\newblock In \emph{2017 IEEE international conference on acoustics, speech and
  signal processing (ICASSP)}, pages 4835--4839. IEEE.

\bibitem[{Kingma and Ba(2015)}]{kingmaadam}
Diederik~P. Kingma and Jimmy Ba. 2015.
\newblock \href {http://arxiv.org/abs/1412.6980} {Adam: {A} method for
  stochastic optimization}.
\newblock In \emph{3rd International Conference on Learning Representations,
  {ICLR} 2015, San Diego, CA, USA, May 7-9, 2015, Conference Track
  Proceedings}.

\bibitem[{Kocabiyikoglu et~al.(2018)Kocabiyikoglu, Besacier, and
  Kraif}]{kocabiyikoglu-etal-2018-augmenting}
Ali~Can Kocabiyikoglu, Laurent Besacier, and Olivier Kraif. 2018.
\newblock \href {https://aclanthology.org/L18-1001} {Augmenting librispeech
  with {F}rench translations: A multimodal corpus for direct speech translation
  evaluation}.
\newblock In \emph{Proceedings of the Eleventh International Conference on
  Language Resources and Evaluation ({LREC} 2018)}, Miyazaki, Japan. European
  Language Resources Association (ELRA).

\bibitem[{Matusov et~al.(2005)Matusov, Ney, and Schlüter}]{matusov2005phrase}
Evgeny Matusov, Hermann Ney, and Ralf Schlüter. 2005.
\newblock Phrase-based translation of speech recognizer word lattices using
  loglinear model combination.
\newblock In \emph{IEEE Workshop on Automatic Speech Recognition and
  Understanding, 2005.}, pages 110--115. IEEE.

\bibitem[{Ney(1999)}]{ney1999speech}
Hermann Ney. 1999.
\newblock Speech translation: Coupling of recognition and translation.
\newblock In \emph{1999 IEEE International Conference on Acoustics, Speech, and
  Signal Processing. Proceedings. ICASSP99 (Cat. No. 99CH36258)}, volume~1,
  pages 517--520. IEEE.

\bibitem[{Ott et~al.(2019)Ott, Edunov, Baevski, Fan, Gross, Ng, Grangier, and
  Auli}]{ott2019fairseq}
Myle Ott, Sergey Edunov, Alexei Baevski, Angela Fan, Sam Gross, Nathan Ng,
  David Grangier, and Michael Auli. 2019.
\newblock fairseq: A fast, extensible toolkit for sequence modeling.
\newblock In \emph{Proceedings of NAACL-HLT 2019: Demonstrations}.

\bibitem[{Park et~al.(2019)Park, Chan, Zhang, Chiu, Zoph, Cubuk, and
  Le}]{park2019specaugment}
Daniel~S Park, William Chan, Yu~Zhang, Chung-Cheng Chiu, Barret Zoph, Ekin~D
  Cubuk, and Quoc~V Le. 2019.
\newblock Specaugment: A simple data augmentation method for automatic speech
  recognition.
\newblock \emph{Proc. Interspeech 2019}, pages 2613--2617.

\bibitem[{Peitz et~al.(2012)Peitz, Wiesler, Nu{\ss}baum-Thom, and
  Ney}]{peitz2012spoken}
Stephan Peitz, Simon Wiesler, Markus Nu{\ss}baum-Thom, and Hermann Ney. 2012.
\newblock Spoken language translation using automatically transcribed text in
  training.
\newblock In \emph{Proceedings of the 9th International Workshop on Spoken
  Language Translation: Papers}.

\bibitem[{Pereyra et~al.(2017)Pereyra, Tucker, Chorowski, Kaiser, and
  Hinton}]{pereyra2017regularizing}
Gabriel Pereyra, George Tucker, Jan Chorowski, {\L}ukasz Kaiser, and Geoffrey
  Hinton. 2017.
\newblock Regularizing neural networks by penalizing confident output
  distributions.
\newblock \emph{5th International Conference on Learning Representations,
  {ICLR} 2017, Toulon, France, April 24-26, 2017, Workshop Track Proceedings}.

\bibitem[{Post et~al.(2013)Post, Kumar, Lopez, Karakos, Callison-Burch, and
  Khudanpur}]{post-etal-2013-improved}
Matt Post, Gaurav Kumar, Adam Lopez, Damianos Karakos, Chris Callison-Burch,
  and Sanjeev Khudanpur. 2013.
\newblock \href {https://aclanthology.org/2013.iwslt-papers.14} {Improved
  speech-to-text translation with the fisher and callhome {S}panish-{E}nglish
  speech translation corpus}.
\newblock In \emph{Proceedings of the 10th International Workshop on Spoken
  Language Translation: Papers}, Heidelberg, Germany.

\bibitem[{Saleem et~al.(2004)Saleem, Jou, Vogel, and Schultz}]{saleem2004using}
Shirin Saleem, Szu-Chen Jou, Stephan Vogel, and Tanja Schultz. 2004.
\newblock Using word lattice information for a tighter coupling in speech
  translation systems.
\newblock In \emph{Proc. Int. Conf. on Spoken Language Processing}, pages
  41--44.

\bibitem[{Sennrich et~al.(2016)Sennrich, Haddow, and
  Birch}]{sennrich2016neural}
Rico Sennrich, Barry Haddow, and Alexandra Birch. 2016.
\newblock Neural machine translation of rare words with subword units.
\newblock In \emph{54th Annual Meeting of the Association for Computational
  Linguistics}, pages 1715--1725. Association for Computational Linguistics
  (ACL).

\bibitem[{Shaw et~al.(2018)Shaw, Uszkoreit, and Vaswani}]{shaw2018self}
Peter Shaw, Jakob Uszkoreit, and Ashish Vaswani. 2018.
\newblock Self-attention with relative position representations.
\newblock In \emph{Proceedings of the 2018 Conference of the North American
  Chapter of the Association for Computational Linguistics: Human Language
  Technologies, Volume 2 (Short Papers)}, pages 464--468.

\bibitem[{Sperber et~al.(2017)Sperber, Neubig, Niehues, and
  Waibel}]{sperber-etal-2017-neural}
Matthias Sperber, Graham Neubig, Jan Niehues, and Alex Waibel. 2017.
\newblock \href {https://doi.org/10.18653/v1/D17-1145} {Neural
  lattice-to-sequence models for uncertain inputs}.
\newblock In \emph{Proceedings of the 2017 Conference on Empirical Methods in
  Natural Language Processing}, pages 1380--1389, Copenhagen, Denmark.
  Association for Computational Linguistics.

\bibitem[{Sperber et~al.(2019)Sperber, Neubig, Niehues, and
  Waibel}]{sperber2019attention}
Matthias Sperber, Graham Neubig, Jan Niehues, and Alex Waibel. 2019.
\newblock Attention-passing models for robust and data-efficient end-to-end
  speech translation.
\newblock \emph{Transactions of the Association for Computational Linguistics},
  7:313--325.

\bibitem[{Sperber and Paulik(2020)}]{sperber-paulik-2020-speech}
Matthias Sperber and Matthias Paulik. 2020.
\newblock \href {https://doi.org/10.18653/v1/2020.acl-main.661} {Speech
  translation and the end-to-end promise: Taking stock of where we are}.
\newblock In \emph{Proceedings of the 58th Annual Meeting of the Association
  for Computational Linguistics}, pages 7409--7421, Online. Association for
  Computational Linguistics.

\bibitem[{Srivastava et~al.(2014)Srivastava, Hinton, Krizhevsky, Sutskever, and
  Salakhutdinov}]{srivastava2014dropout}
Nitish Srivastava, Geoffrey Hinton, Alex Krizhevsky, Ilya Sutskever, and Ruslan
  Salakhutdinov. 2014.
\newblock Dropout: a simple way to prevent neural networks from overfitting.
\newblock \emph{The journal of machine learning research}, 15(1):1929--1958.

\bibitem[{Stentiford and Steer(1988)}]{stentiford1988machine}
Fred~WM Stentiford and Martin~G Steer. 1988.
\newblock Machine translation of speech.
\newblock \emph{British Telecom technology journal}, 6(2):116--122.

\bibitem[{Vaswani et~al.(2017)Vaswani, Shazeer, Parmar, Uszkoreit, Jones,
  Gomez, Kaiser, and Polosukhin}]{vaswani2017attention}
Ashish Vaswani, Noam Shazeer, Niki Parmar, Jakob Uszkoreit, Llion Jones,
  Aidan~N Gomez, {\L}ukasz Kaiser, and Illia Polosukhin. 2017.
\newblock \href {http://arxiv.org/abs/1706.03762} {Attention is all you need}.
\newblock \emph{Advances in neural information processing systems}, 30.

\bibitem[{Waibel et~al.(1991)Waibel, Jain, McNair, Saito, Hauptmann, and
  Tebelskis}]{waibel1991janus}
Alex Waibel, Ajay~N Jain, Arthur~E McNair, Hiroaki Saito, Alexander~G
  Hauptmann, and Joe Tebelskis. 1991.
\newblock Janus: a speech-to-speech translation system using connectionist and
  symbolic processing strategies.
\newblock In \emph{Acoustics, Speech, and Signal Processing, IEEE International
  Conference on}, pages 793--796. IEEE Computer Society.

\bibitem[{Wang et~al.(2020)Wang, Wu, Liu, Yang, and Zhou}]{wang2020bridging}
Chengyi Wang, Yu~Wu, Shujie Liu, Zhenglu Yang, and Ming Zhou. 2020.
\newblock Bridging the gap between pre-training and fine-tuning for end-to-end
  speech translation.
\newblock In \emph{Proceedings of the AAAI Conference on Artificial
  Intelligence}, pages 9161--9168.

\bibitem[{Xu et~al.(2021)Xu, Hu, Li, Zhang, Huang, Ju, Xiao, and
  Zhu}]{xu2021stacked}
Chen Xu, Bojie Hu, Yanyang Li, Yuhao Zhang, Shen Huang, Qi~Ju, Tong Xiao, and
  Jingbo Zhu. 2021.
\newblock Stacked acoustic-and-textual encoding: Integrating the pre-trained
  models into speech translation encoders.
\newblock In \emph{Proceedings of the 59th Annual Meeting of the Association
  for Computational Linguistics and the 11th International Joint Conference on
  Natural Language Processing (Volume 1: Long Papers)}, pages 2619--2630.

\bibitem[{Zhang et~al.(2019)Zhang, Ge, Chen, and Fan}]{zhang-etal-2019-lattice}
Pei Zhang, Niyu Ge, Boxing Chen, and Kai Fan. 2019.
\newblock \href {https://doi.org/10.18653/v1/P19-1649} {Lattice transformer for
  speech translation}.
\newblock In \emph{Proceedings of the 57th Annual Meeting of the Association
  for Computational Linguistics}, pages 6475--6484, Florence, Italy.
  Association for Computational Linguistics.

\end{thebibliography}
\bibliographystyle{acl_natbib}

\newpage
\appendix

\section{Experimental Setup}
\label{sec:appendix_setup}
For the ASR data, we extract 80-dimensional MFCC features every 10ms. 
The text data is post-processed by lower-casing, removing punctuation and transcriber tags, and by applying BPE~\cite{sennrich2016neural} e a vocabulary size of 8000.
We post-process the source text to be transcript-like, by removing punctuation and lower-casing. On both source and target sentences, we apply BPE with a vocabulary size of 32k. 
For models using tight integration, a separate translation model is trained using the transcription vocabulary.
In both cases, the validation set is the concatenation of the validation sets provided by MuST-C and IWSLT TED.

Our implementation is based on fairseq~\cite{ott2019fairseq} and is available online\footnote{\url{https://github.com/tran-khoa/joint-training-cascaded-st}}.

\paragraph{ASR model} 
For speech recognition, we use a Transformer model \cite{vaswani2017attention} with 12 encoder layers and 6 decoder layers. 
Instead of absolute positional encodings, we use relative positional encodings as introduced by \cite{shaw2018self}. 
We reduce the audio feature length using a two-layer convolutional network with kernel size 5 and 1024 channels. Other parameters are adapted from the original base configuration. Our ASR model consists of 70M trained parameters.

Each epoch is split into 20 sub-epochs. 
We use the Adam optimizer~\cite{kingmaadam} with learning rate 0.0003 and 10 warmup sub-epochs, and the learning rate is scaled by 0.8 for every 3 sub-epochs without improvement on the validation set. 
As regularization, we use SpecAugment~\cite{park2019specaugment} ($(F, m_F, T, m_T,p , W)=(16,4,40,2,1.0,0)$), a dropout~\cite{srivastava2014dropout} of 0.1, and label smoothing~\cite{pereyra2017regularizing} with $\alpha=0.1$. Additionally, we train a CTC loss as an additional task during training \cite{kim2017joint}.
\paragraph{MT model} For translation, we also use a Transformer model with 6 encoder and 6 decoder layers. Again, we use relative positional encodings, other parameters are adapted from the original base configuration, resulting in a model with 70M trained parameters.

We use the same optimization procedure as with the ASR model, except that we now start with 4000 warmup steps. 
As regularization, we use a dropout of 0.1 and label smoothing with $\alpha=0.1$.

\paragraph{Joint Training} We use the same training setup as for the ASR model, but with a learning rate of $3\times 10^{-6}$. Furthermore, as we are now working on a smaller ST dataset, an epoch is split into 10 sub-epochs.

\paragraph{Fine-tuning} We fine-tune the ASR model on all in-domain ASR data (600K segments) with a learning rate of 0.0001, while the MT model is fine-tuned on the MuST-C dataset (300K parallel sentences) with a learning rate of 0.00001.

We use beam search with $N=12$.
All BLEU scores reported are calculated on case-sensitive data using the official WMT scoring script.
TER scores are calculated using TERCom.
For top-$K$ experiments, we use $K=4$.

We train our experiments on NVIDIA GTX 1080 Ti. Training the ASR takes around 4 weeks, all other experiment take around 2-3 weeks. The average runtime for inference on non-joint experiments is about 15 minutes, where joint experiments need around 2 hours.

\end{document}